\documentclass[showpacs,amsmath,superscriptaddress,reprint,aps]{revtex4-1}
\usepackage{graphicx}
\usepackage{hyperref}

\begin{document}

\title
{Super-Resolution Fluorescence Imaging of Carbon Nanotubes Using a Nonlinear Excitonic Process}
\author{K.~Otsuka}
\affiliation{Quantum Optoelectronics Research Team, RIKEN Center for Advanced Photonics, Saitama 351-0198, Japan}
\affiliation{Nanoscale Quantum Photonics Laboratory, RIKEN Cluster for Pioneering Research, Saitama 351-0198, Japan}
\author{A.~Ishii}
\affiliation{Quantum Optoelectronics Research Team, RIKEN Center for Advanced Photonics, Saitama 351-0198, Japan}
\affiliation{Nanoscale Quantum Photonics Laboratory, RIKEN Cluster for Pioneering Research, Saitama 351-0198, Japan}
\author{Y.~K.~Kato}
\email[Corresponding author. ]{yuichiro.kato@riken.jp}
\affiliation{Quantum Optoelectronics Research Team, RIKEN Center for Advanced Photonics, Saitama 351-0198, Japan}
\affiliation{Nanoscale Quantum Photonics Laboratory, RIKEN Cluster for Pioneering Research, Saitama 351-0198, Japan}

\begin{abstract}
Highly efficient exciton-exciton annihilation process unique to one-dimensional systems is utilized for super-resolution imaging of air-suspended carbon nanotubes. Through the comparison of fluorescence signals in linear and sublinear regimes at different excitation powers, we extract the efficiency of the annihilation processes using conventional confocal microscopy. Spatial images of the annihilation rate of the excitons have resolution beyond the diffraction limit. We investigate excitation power dependence of the annihilation processes by experiment and Monte Carlo simulation, and the resolution improvement of the annihilation images can be quantitatively explained by the superlinearity of the annihilation process. We have also developed another method in which the cubic dependence of the annihilation rate on exciton density is utilized to achieve further sharpening of single nanotube images. 
\end{abstract}

\maketitle

As a result of strong Coulomb interaction arising from the one-dimensional (1D) nature of single-walled carbon nanotubes (CNTs), electron-hole pairs form excitons that are stable even at room temperature~\cite{Ogawa:1991, Ando:1997, Wang:2005}. Confinement and diffusion~\cite{Murakami:2009prb2, Moritsubo:2010, Yoshikawa:2010, Xie:2012} of the excitons in a nanotube lead to their efficient annihilation process upon collision with one another~\cite{Wang:2004prb, Ma:2005, Xiao:2010}, resulting in a peculiar cubic dependence of the exciton-exciton annihilation (EEA) rate on the density of excitons~\cite{Ishii:2015, Khasminskaya:2016}. The efficient EEA can be, for example, utilized for room-temperature single photon generation at telecommunication wavelengths~\cite{Ma:2015prl, Ishii:2017}. The diameter-dependent wavelength of nanotube fluorescence also includes the near-infrared window, where scattering is small and absorption is weak, allowing for deep-tissue imaging using CNTs as fluorescent agents~\cite{Hong:2012, Yomogida:2016, Aota:2016, Danne:2017}. As advanced techniques for super-resolution imaging~\cite{Rust:2006, Cognet:2008, Hell:2015, Danne:2018}, such as two-photon excitation microscopy and stimulated emission depletion microscopy, rely on the nonlinear optical response in fluorescence agents~\cite{Denk:1990, Hell:1994, Heintzmann:2002, Fujita:2007, Chmyrov:2013}, the EEA process could play a key role in the development of nanotube-based biological imaging as well.

Here we demonstrate subdiffraction imaging of air-suspended CNTs by extracting the nonlinear EEA component using a typical confocal microscopy system. By combining two fluorescence images obtained at different excitation powers, an EEA rate image with enhanced resolution can be constructed. Excitation power-dependence of the extracted EEA efficiency and the spatial resolution of the EEA imaging are experimentally investigated, and we perform Monte Carlo simulation of the EEA process to identify the resolution limit of this technique. In addition to the use of nonlinearity between the EEA rate and the exciton generation rate, the cubic dependence of the EEA rate on exciton density is utilized in another protocol for super-resolution imaging of CNTs. By measuring the excitation power required to establish a predefined exciton density, we are able to achieve even narrower widths for isolated nanotube images.

\section*{Results and discussion}
\begin{figure}
\includegraphics{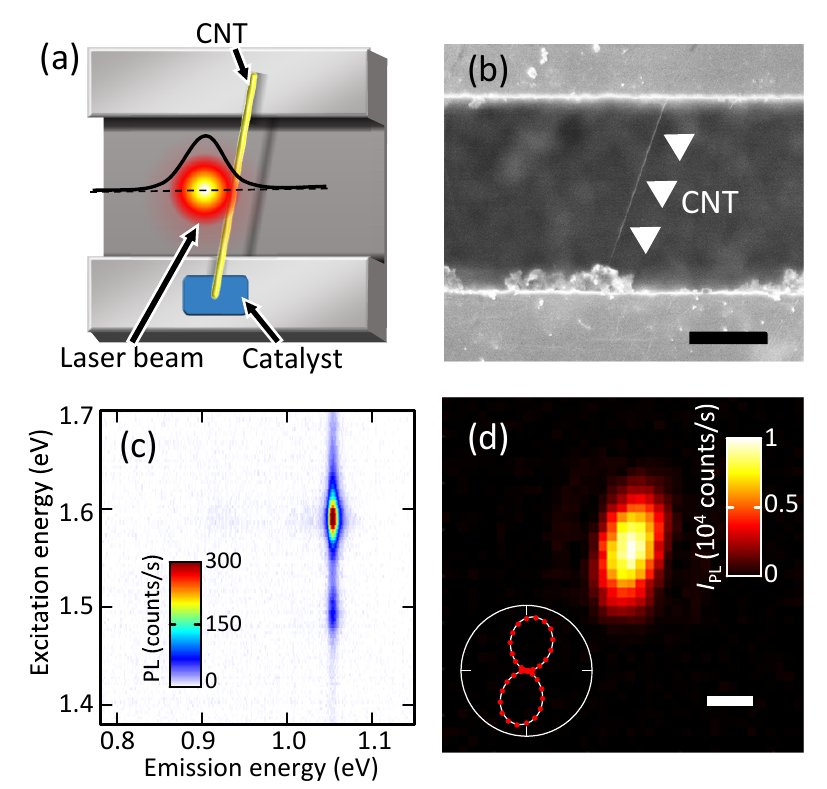}
\caption{
\label{Fig1} (a) A schematic of an air-suspended nanotube sample. For optical imaging, the samples are scanned along trenches relative to the fixed laser beam. (b) A scanning electron micrograph of a typical nanotube. (c) A PL excitation map for a (11,3) nanotube. The excitation power is 0.1~$\mu$W. (d) A PL image of the nanotube measured in (c). Inset: Polarization dependence of $I_\mathrm{PL}$. Excitation power and energy used for imaging are 0.1~$\mu$W and $E_{22}=1.590$~eV, respectively, and the image is extracted at an emission energy of $E_{11}=1.055$~eV with a spectral integration window of $\sim$50~meV. Scale bars in (b) and (d) are 500~nm.
}
\end{figure}

\paragraph*{Excitation Power Dependence.}
Our samples are as-grown nanotubes suspended over trenches on silicon substrates. A schematic and a scanning electron micrograph of an air-suspended nanotube are shown in Figs.~\ref{Fig1}(a) and \ref{Fig1}(b), respectively. We perform photoluminescence (PL) measurements on such samples using a homebuilt confocal microscopy system~\cite{Ishii:2015}, in which excitation laser power is controlled in a wide range by a continuously variable neutral density filter.

Figure~\ref{Fig1}(c) shows PL excitation spectroscopy data performed on a single air-suspended nanotube with a length $L\approx1.2~\mu$m, and we determine the tube chirality to be (11,3). We then investigate laser polarization dependence of PL intensity with excitation energy at the $E_{22}$ resonance [inset of Fig.~\ref{Fig1}(d)]. For the following PL measurements, the polarization angle and the excitation energy are fixed parallel to the tube axis and at the $E_{22}$ resonance, respectively. Figure 1(d) shows an image of the integrated PL intensity $I_\mathrm{PL}$ obtained with a $\sim$50-meV-wide spectral integration window centered at the $E_{11}$ resonance. The width of the nanotube PL image is predominantly determined by the excitation laser beam profile, and thus much larger than the actual nanotube diameter of 1.0~nm.

\begin{figure}
\includegraphics{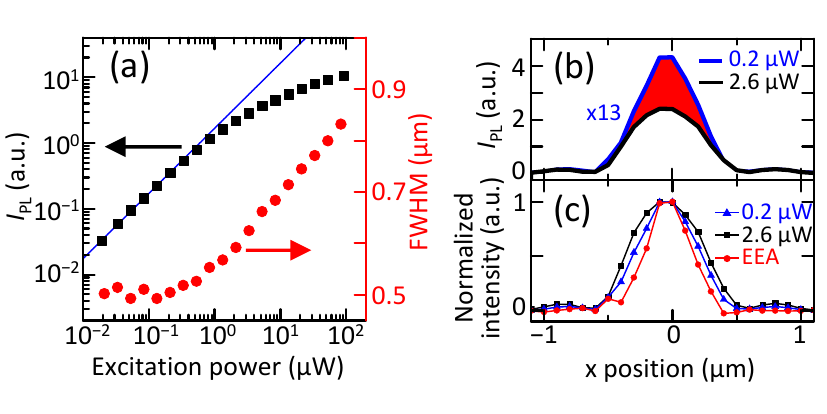}
\caption{
\label{Fig2} (a) Excitation power dependence of PL intensity $I_\mathrm{PL}$ with the laser focused at the center of the nanotube (black squares) and FWHM of PL intensity profiles along the trench direction (red circles). The nanotube is the same one as in Fig.~\ref{Fig1}(c) and \ref{Fig1}(d). $I_\mathrm{PL}$ is obtained by calculating the peak area of a Lorentzian fit to the emission spectrum. (b) PL intensity profiles scaled by excitation powers. Difference of the black and the blue lines is proportional to the EEA rate $\Gamma_\mathrm{EEA}$. (c) Normalized profiles of $I_\mathrm{PL}$ for the two different excitation powers and the EEA rate obtained from the subtraction in (b). 
}
\end{figure}

The resolution enhancement will be achieved by using the efficient EEA process, whose effects can be observed in the power dependence of PL intensity [Fig.~\ref{Fig2}(a), black squares]. $I_\mathrm{PL}$ is proportional to the excitation power at low powers, but shows a cubic root dependence  at high powers due to EEA~\cite{Ishii:2015}. The blue line in Fig.~\ref{Fig2}(a) is a linear fit to the low-power results, corresponding to the PL intensity expected in the absence of EEA. The deviation of actual PL intensity from the blue line would correspond to the EEA rate $\Gamma_\mathrm{EEA}$. 

The efficient EEA process also affects the imaging resolution. We measure 1D $I_\mathrm{PL}$ profiles of the nanotube along the trench direction with various excitation powers, and the full-width at half-maximum (FWHM) $w_\mathrm{h}$ of the profiles are plotted as red circles in Fig.~\ref{Fig2}(a) as a function of the excitation power. At the low power limit, PL intensity profiles reflect the excitation laser beam profile because $I_\mathrm{PL}$ is proportional to the exciton generation rate $g$ in the absence of EEA. At the other extreme where $I_\mathrm{PL}$ is proportional to $g^{1/3}$ due to the cubic-law EEA process, we expect the effects of nonlinearity on the width of the PL intensity profile. If we approximate the laser beam profile by a Gaussian function $\exp (-2x^2/r^2)$ with $r$ being the laser 1/$e^2$ radius, the intensity profile of PL that has $g^\alpha$ dependence becomes $I_\mathrm{PL}(x) \propto \exp (-2\alpha x^2/r^2)=\exp [-2x^2/(r/\sqrt{\alpha})^2]$, where $\alpha$ is the power exponent of the generation rate dependence. The FWHM changes by a factor
\begin{equation}
\frac{r/\sqrt{\alpha}\sqrt{2\ln 2}}{r\sqrt{2\ln 2}}= \frac{1}{\sqrt{\alpha}},
\label{DiffEq1}
\end{equation}
and it is reasonable that the width of $I_\mathrm{PL}$ profile increases by a factor of $\sqrt{3}$ at the highest power.

\paragraph*{Extracting the Influence of EEA.}
Although it may seem as if the EEA process has a negative effect on resolution, we can utilize the nonlinear power dependence on the generation rate to achieve enhanced spatial resolution. We extract the EEA component from PL intensities measured at two different excitation powers, where the EEA extraction power $P_1$ is higher than the reference power $P_2$. Figure~\ref{Fig2}(b) illustrates how the EEA component is extracted. The black line is the PL intensity profile obtained with the EEA extraction power $P_1=2.6$~$\mu$W, showing slight broadening caused by EEA. The blue line is the intensity profile at $P_2=0.2$~$\mu$W scaled by the ratio of the excitation powers, which is equivalent to the expected PL intensity profile in the absence of the EEA process. The difference between the black and blue lines (red region) is proportional to the EEA rate $\Gamma_\mathrm{EEA}$ in the nanotube.

Figure~\ref{Fig2}(c) shows normalized profiles of raw PL intensity (black and blue for $P_1$ and $P_2$, respectively) and the EEA component extracted by the protocol described above (red). The width of the intensity profile at the excitation power of 0.2~$\mu$W corresponds to the minimum width obtained in the power-dependent width in Fig.~\ref{Fig2}(a), but the profile of the extracted EEA component has an even smaller width. As $\Gamma_\mathrm{EEA}$ increases superlinearly with $g$, the spatial profile of the EEA component peaks sharply when the laser is centered on the nanotube.

\begin{figure}
\includegraphics{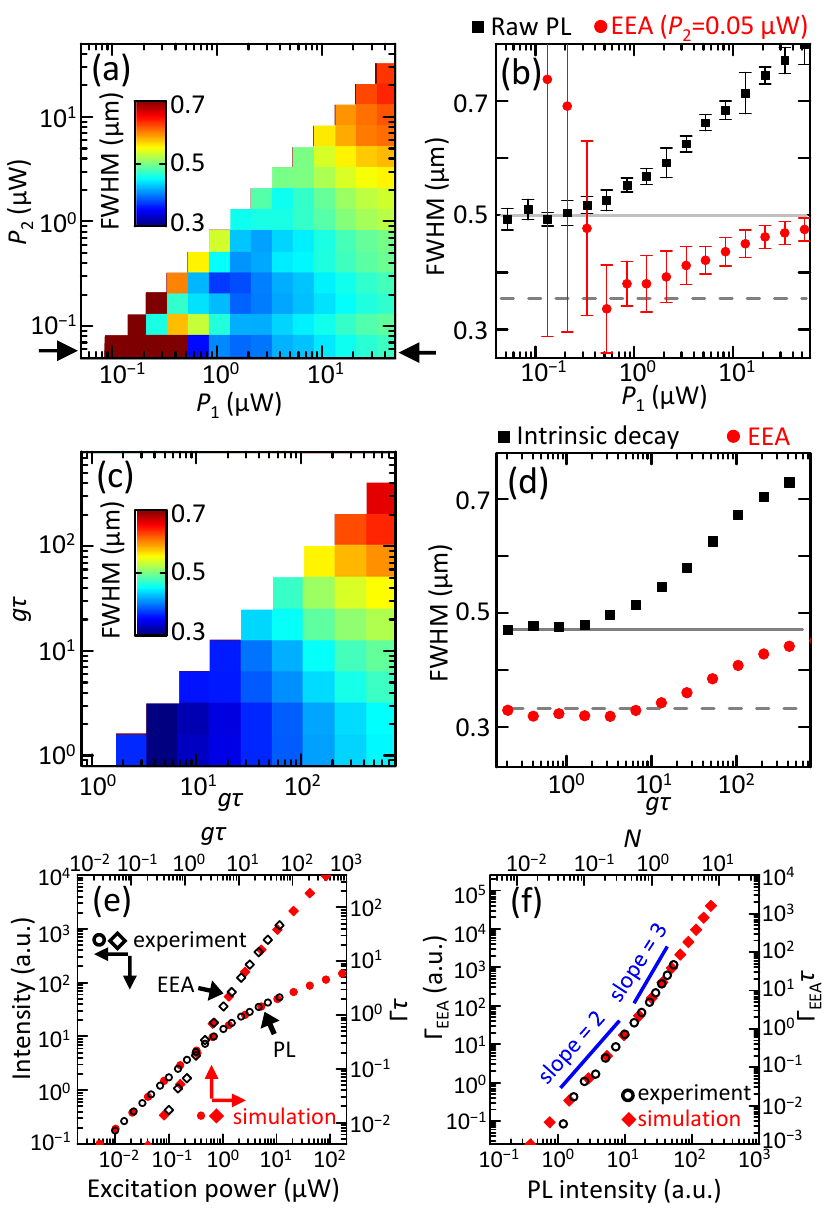}
\caption{
\label{Fig3} (a) FWHM of 1D EEA profiles as a function of the EEA extraction power $P_1$ and the reference power $P_2$ used for the subtraction of $I_\mathrm{PL}$. The data are averaged for seven measurements repeated with the same condition. (b) FWHM of $I_\mathrm{PL}$ and the extracted EEA rate profiles for a fixed power of $P_2=0.05$~$\mu$W. Error bars represent the standard deviation. Solid and broken gray lines correspond to the laser beam FWHM and that multiplied by $1/\sqrt{2}$, respectively. (c) FWHM as a function of the EEA extraction and the reference generation rate from the Monte Carlo simulations. (d) FWHM of the intrinsic decay rate profiles and the EEA rate profiles from the simulations. (e) Excitation power dependence of the PL intensity and the extracted EEA component for the experiments and the simulations. (f) EEA rates as a function of $I_\mathrm{PL}$ and $N$ from the experiments and the simulations, respectively.
}
\end{figure}

To characterize the resolution and signal-to-noise ratio, we perform experiments using a range of excitation power combinations for extracting the EEA rate $\Gamma_\mathrm{EEA}$. Measurements are repeated 7 times for each combination of the EEA extraction power $P_1$ and the reference power $P_2$, and the average width of the $\Gamma_\mathrm{EEA}$ profiles is plotted in Fig.~\ref{Fig3}(a) as a function of $P_1$ and $P_2$. Roughly speaking, large $P_1$ and $P_2$ result in large $w_\mathrm{h}$ of the $\Gamma_\mathrm{EEA}$ profiles, while low powers  are preferable for high-resolution imaging. 

We consider the widths obtained at the lowest $P_2=0.05$~$\mu$W to quantitatively evaluate the resolution improvement. Figure~\ref{Fig3}(b) shows the $P_1$ dependence of $w_\mathrm{h}$ obtained from the raw PL intensity profiles (black squares) and the extracted EEA profiles with $P_2$ fixed at 0.05~$\mu$W (red circles). Error bars represent the standard deviation for the repeated measurements. The FWHM from raw PL reproduces the behavior observed in Fig.~\ref{Fig2}(a), and the error values are similar over the entire range of the excitation power. In comparison, the width for EEA rate profiles decreases from $\sim$500 to $\sim$350~nm as $P_1$ decreases, which is approximately an improvement by a factor of $\sqrt{2}$. We note that the width of the EEA profiles has large error bars at low powers because of the small $\Gamma_\mathrm{EEA}$ signals that depends superlinearly on $g$.

\paragraph*{Monte Carlo Simulation.}
The minimum width achievable from the EEA profile in an ideal situation with negligible noise is investigated by conducting Monte Carlo simulation~\cite{Mouri:2014, Ishii:2015} using parameters directly comparable with the experimental results. Excitonic processes, such as exciton generation, diffusion, and decay, are stochastically evaluated. Excitation profile has a spatial distribution $\exp [-2(x^2+y^2)/r^2]$, where $x$ and $y$ are positions perpendicular to and along the nanotube, respectively.
1D rate profiles are calculated by moving the nanotube position in the $x$ direction, simulating the experiment performed to obtain the $I_\mathrm{PL}$ profiles shown in Fig.~\ref{Fig2}(b). We compute the intrinsic decay rate $\Gamma_\mathrm{I}$ from the time-averaged number of excitons that go through the intrinsic decay, which corresponds to the PL intensity in the experiments. The simulations are repeated for various photon incident rates, and the EEA rate profiles are computed from $\Gamma_\mathrm{I}$ profiles in a manner similar to the measurements for Fig.~\ref{Fig3}(a). Figure~\ref{Fig3}(c) shows the FWHM of the extracted EEA rate profiles as a function of the EEA extraction and reference generation rates. Data are plotted in terms of a unitless parameter $g\tau$, representing the average number of excitons that are generated during the exciton intrinsic decay lifetime $\tau$. Note that we use the exciton generation rate evaluated with the nanotube at the center of the laser beam in Fig.~\ref{Fig3}, in which case it becomes equivalent to the excitation power in the experiments. The simulation results exhibit a trend similar to the experiments; large $w_\mathrm{h}$ for combinations of large $g\tau$ and {\it vice versa}. 

Using the simulation data, it is possible to directly evaluate the EEA rates $\Gamma_\mathrm{EEA}$. In Fig.~\ref{Fig3}(d), the FWHM of the EEA profiles as well as the intrinsic decay profiles are plotted, and both reproduce the overall behavior of the experimental data. As the generation rate is decreased, the width of the intrinsic decay approaches that of the simulated laser profile as indicated by the solid gray line. The EEA profile width, in comparison, is already below the solid line at large generation rates. The width decreases with the generation rate, and approaches the broken gray line which corresponds to 1/$\sqrt{2}$ width of the laser profile. According to Eq.~(\ref{DiffEq1}),  the change in $w_\mathrm{h}$ implies that the EEA rate has a nontrivial dependence on the exciton generation rate. 

\paragraph*{EEA Rate vs Generation Rate.}
We experimentally confirm the $g$ dependence of $\Gamma_\mathrm{EEA}$ by measuring the PL intensity [Fig.~\ref{Fig3}(e), open circles] and subtracting a linear fit to the low power PL data (open diamonds). Similarly, the $g\tau$ dependence of the intrinsic decay rate (filled circles) and the EEA rate (filled diamonds) obtained from the simulation is overlaid in Fig.~\ref{Fig3}(e) in terms of unitless parameters $\Gamma_\mathrm{I} \tau$ and $\Gamma_\mathrm{EEA} \tau$, respectively. The simulation is in good agreement with the experiment, showing that our simple model accurately describes the behavior of excitons in nanotubes. In this log-log plot, $\Gamma_\mathrm{I}$ has slopes of 1 and 1/3 at low and high generation rates, respectively, as also observed in Fig.~\ref{Fig2}(a). The slope of $\Gamma_\mathrm{EEA}$ is of our interest, which is 2 at low powers and approaches 1 as the power is increased. Through Eq.~(\ref{DiffEq1}), the observed values of the slope can quantitatively explain the power dependence of $w_\mathrm{h}$ obtained from the EEA profiles [Fig.~\ref{Fig3}(d)]. 

The $\Gamma_\mathrm{EEA}$ slope of 1 at high power is reasonable, because $g\approx \Gamma_\mathrm{EEA}$ with EEA being the dominant decay process. The quadratic dependence of $\Gamma_\mathrm{EEA}$, however, seems to contradict with the cubic dependence expected for EEA~\cite{Ishii:2015}. In order to resolve this apparent inconsistency, $\Gamma_\mathrm{EEA}\tau$ (filled diamonds) is plotted as a function of time-averaged exciton number $N=\Gamma_\mathrm{I}\tau$ [Fig. 3(f)], and we find a transition from the quadratic to cubic dependence near $N=1$. Similarly, the experimentally extracted EEA rate is replotted as a function of $I_\mathrm{PL}$ (open diamonds) in Fig.~\ref{Fig3}(f), which coincides with the simulation result. The transition in the $N$ dependence of the EEA rate can be explained by the fact that EEA process occurs only when multiple excitons coexist in a nanotube. When $N>1$, $\Gamma_\mathrm{EEA}$ in a 1D system is proportional to $N^3$ as discussed previously~\cite{Ishii:2015}. When the exciton number $N\ll 1$, however, the situation changes. The probability for the instantaneous exciton number $\lambda \geq3$ is negligible, while EEA does not occur for $\lambda =1$. The EEA rate is then dominated by the case where two excitons coexist, whose probability is given by the Poisson distribution $p(\lambda) = N^\lambda e^{-N}/ \lambda !$ to be $p(2)\approx N^2/2$. In this regime, multiple exciton generation, rather than exciton diffusion, is the limiting factor of the EEA process. 

We can further obtain an explicit expression for the EEA rate. After time $t$ from the first exciton generation, the survival probability of the first exciton and the arrival probability of the second exciton are $p_\mathrm{s}(t)=\exp (-t/\tau)$ and $p_\mathrm{a}(t)=gt\exp (-gt)$, respectively. The expected value of the arrival time interval $\int t p_\mathrm{s}(t)p_\mathrm{a}(t) dt/\int p_\mathrm{s}(t) p_\mathrm{a}(t) dt\approx 2\tau$ of coexisting excitons is independent of $N$ when $N \ll 1$. The initial distance between the two excitons after diffusion is therefore constant ($\sim \sqrt{8D\tau/\pi}$) with $D$ being the diffusion constant. Despite the diffusion of the first exciton, the collision rate of such exciton pairs is then $\sim \pi /2\tau$ under the condition of $\lambda =2$, and $\Gamma_\mathrm{EEA} = p(2) \times \pi/2\tau = \pi N^2/4\tau$.

\paragraph*{2D EEA Imaging.}
Having understood the mechanism for the resolution improvement, we perform two-dimensional (2D) imaging of the raw PL intensity and the EEA rate. According to the analysis performed in Figs.~\ref{Fig3}(a) and \ref{Fig3}(c), the reference power $P_2$ should be small enough to avoid EEA while the EEA extraction power $P_1$ also should be in the linear regime of $I_\mathrm{PL}$ vs the excitation power in order to obtain high resolution images. We note, however, that $P_1$ should not be too close to $P_2$ to keep the signal-to-noise ratio sufficiently high. We thus choose $P_1=0.8$~$\mu$W and $P_2=0.1$~$\mu$W, where the excitation power is modulated by switching the neutral density filters at every step of the sample scan.

\begin{figure}
\includegraphics{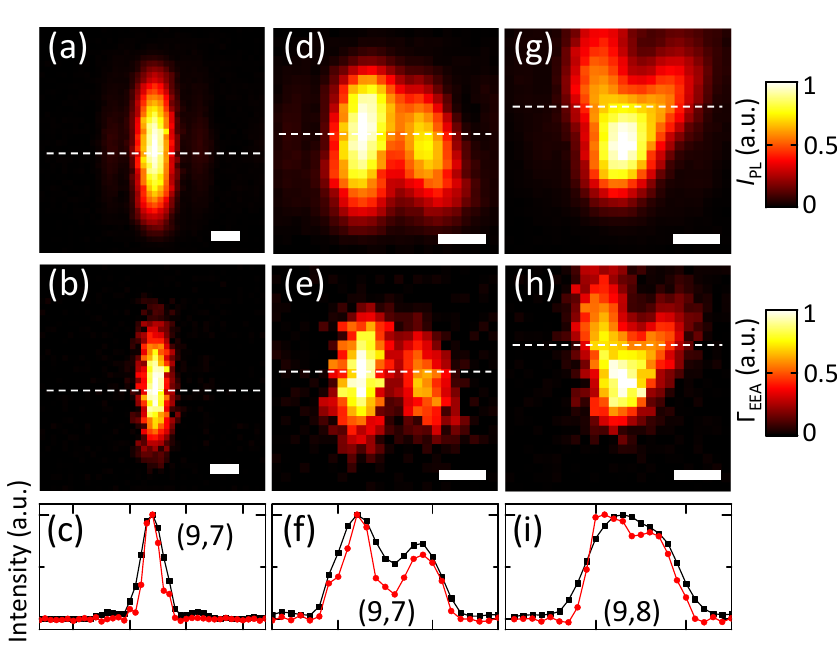}
\caption{
\label{Fig4} 2D images of a (9,7) nanotube for (a) the PL intensity and (b) the extracted EEA rate. (c) 1D profiles from the PL (black) and EEA (red) images at the same position indicated by broken white lines. (d--i) Similar sets of 2D images and the 1D profiles for the two nanotubes with (d--f) a parallel and (g--i) a Y-shaped configuration. The excitation energies are fixed at the $E_{22}$ of each nanotube, and all the images were extracted at the $E_{11}$ energy with a spectral window of $\sim$50~meV. All the images are normalized by their maximum intensity so that the image resolution can be easily compared. Scale bars are 500~nm.
}
\end{figure}

Figure~\ref{Fig4} displays the optical images for three configurations of nanotubes. Because $P_2=0.1$~$\mu$W used for Figs.~\ref{Fig4}(a,d,g) is in the linear regime of $I_\mathrm{PL}$, the width of the raw PL image is solely limited by the laser beam profile and cannot be further reduced by lowering the power. In the simplest case of a single nanotube, the EEA image of the (9,7) nanotube [Fig.~\ref{Fig4}(b)] has even smaller width than the PL image [Fig.~\ref{Fig4}(a)]. We note that the EEA image represents the degree of PL intensity reduction through the EEA process. When two (9,7) tubes are lying closely, the tubes are more clearly resolved by extracting the EEA component, which is also apparent from the 1D profiles shown in Fig.~\ref{Fig4}(f). For a Y-shaped junction of (9,8) nanotubes, we show 1D profiles at the position where the $I_\mathrm{PL}$ profile of the two tubes cannot be resolved [black line in Fig.~\ref{Fig4}(i)]. These nanotubes are successfully separated through the super-resolution imaging of EEA rates. It is noteworthy that the nonlinearity in nanotubes enables subdiffraction imaging at a power density as low as $\sim$300~W/cm$^2$ with a continuous wave laser. This value is, for example, two-orders of magnitude smaller than that used in the high-resolution microscopy which uses fluorescence saturation of Rhodamine 6G~\cite{Fujita:2007}, benefiting from the highly mobile excitons confined in 1D~\cite{Xiao:2010}. 

\begin{figure}
\includegraphics{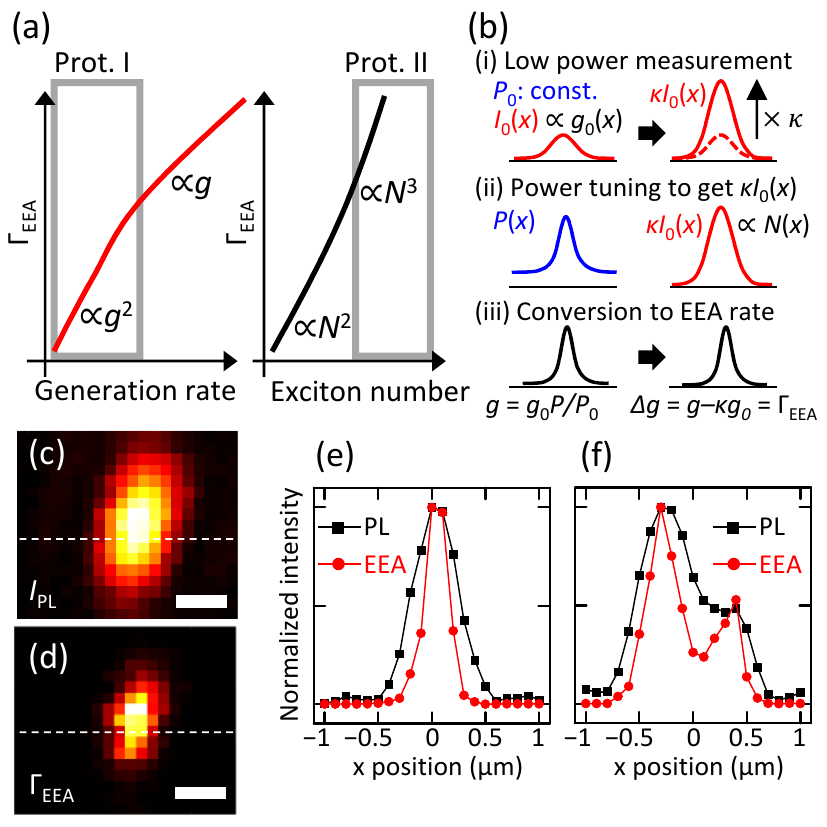}
\caption{
\label{Fig5} (a) Illustration for utilizing the nonlinearity of $\Gamma_\mathrm{EEA}$ as a function of the exciton generation rate $g$ or the exciton number $N$. While quadratic nonlinearity of $\Gamma_\mathrm{EEA}$ at small $g$ is used for Protocol I (used in Figs.~\ref{Fig1}--\ref{Fig4}), cubic nonlinearity of $\Gamma_\mathrm{EEA}$ against $N$ is used in protocol II. (b) Schematics of the protocol to extract the cubic nonlinearity. (c) PL and (d) EEA images of the (11,3) nanotube with $P_0=0.2~\mu$W and $\kappa=25$. Scale bars are 500~nm. (e) 1D intensity profiles of the nanotube shown in (c) and (d). (f) Intensity profiles of the nanotubes shown in Figs.~\ref{Fig4}(d--f). 
}
\end{figure}

\paragraph*{Cubic-Law EEA at High Powers.}
We now consider whether it is possible to utilize the cubic dependence of the EEA rate $\Gamma_\mathrm{EEA}$ on $N$ to achieve even higher resolution images. In the imaging protocol employed in Figs.~\ref{Fig2}--\ref{Fig4}, the $g^2$ dependence of $\Gamma_\mathrm{EEA}$ is used [Protocol I, left panel of Fig.~\ref{Fig5}(a)]. In comparison, we introduce a protocol that exploits the $N^3$ term of the EEA rate that appears at large $N$ [Protocol II, right panel of Fig.~\ref{Fig5}(a)]. Instead of laser position dependent exciton generation rate $g(x)$ resulting from a constant excitation power in Protocol I, we create laser position dependent exciton number $N(x)$ that reproduces the laser beam profile. The EEA rate extracted from such $N(x)$ should allow for resolution enhancement through the cubic dependence. As briefly illustrated in Fig.~\ref{Fig5}(b), we first obtain a PL intensity profile $I_0(x)$ at a sufficiently low excitation power $P_0$ to purely extract the generation rate profile $g_0(x)$. Then $I_0(x)$ is multiplied by a constant $\kappa>1$, which will be the target PL intensity $\kappa I_0(x)$ for the next scan. We reproduce the target PL intensity profile $\kappa I_0(x)$ by controlling the excitation power $P(x)$ to compensate for the intensity loss due to EEA. $P(x)$ is tuned at every point until the intensity difference between the measurement and $\kappa I_0(x)$ becomes less than 3\%. The position-dependent exciton generation rate for this scan $g(x)=g_0(x)P(x)/P_0$ is used to extract the EEA rate $\Gamma_\mathrm{EEA}(x)$. As the EEA rate is equal to the additional exciton generation rate compared with that expected in the absence of EEA, $\Gamma_\mathrm{EEA}(x)=\Delta g(x)=g(x)-\kappa g_0(x)$.
 
Figures~\ref{Fig5}(c) and \ref{Fig5}(d) show the raw PL image and the EEA rate image, respectively, obtained through Protocol II. Since larger $\kappa$ gives smaller $w_\mathrm{h}$, we choose $\kappa=25$ and $P_0=0.2$~$\mu$W to fully utilize the $N^3$ nonlinearity at high powers, resulting in the maximum $P(x)=53$~$\mu$W. Cross sectional profiles of these images are shown in Fig.~\ref{Fig5}(e). Clearly, the EEA rate image shows a narrower profile. If we take horizontal cross sections of the images and average the width along the length of the tube, the EEA image gives an average $w_\mathrm{h}$ of 290~nm, while that for the PL image is 520~nm. The width reduction corresponds to a factor of $\sqrt{3}$ as expected from the $N^3$ dependence. As shown in Fig.~\ref{Fig5}(f), this method can resolve multiple CNTs lying closely as a result of the improved spatial resolution. It should be noted, however, that the spatial resolution of the image does not always improve as much as the width	 observed for single tubes. Unlike the case in Protocol I, where the linear component of the PL is canceled out during the subtraction, the linear component from adjacent nanotubes is included in the signal during the power tuning step of Protocol II. The inclusion of the linear component inevitably causes a reduction of the nonlinear EEA component, resulting in a suboptimal resolution for high excitation powers.

\section*{Conclusions}
In summary, we have performed super-resolution imaging of air-suspended CNTs by visualizing the efficient EEA process using two protocols compatible with typical confocal microscopy systems. By subtracting the linear PL component to extract the nonlinear EEA rate, the spatial resolution improves by a factor of $\sqrt{2}$ compared to the diffraction limit. The resolution improvement is determined by the $N^2$ dependent EEA rate at low exciton numbers, as confirmed by Monte Carlo simulations. To utilize the $N^3$ dependence unique to the 1D system of nanotubes, we have developed another protocol for super-resolution imaging in which the excitation power is adaptively tuned during the measurement. Using the second protocol, the width of a single nanotube can be as narrow as $1/\sqrt{3}$ compared to the conventional PL imaging. 

\section*{Methods}
\paragraph*{Air-Suspended Carbon Nanotubes.}
The air-suspended nanotubes are synthesized by alcohol chemical vapor deposition~\cite{Maruyama:2002, Ishii:2015}. Trenches are formed on Si substrates through electron beam lithography and dry etching. We use Fe(III) acetylacetonate and fumed silica in ethanol as catalyst, where spin-coating and lift-off processes are used to deposit the catalyst in a region defined by another lithography process. After heating in air at 400$^\circ$C for 5~min, CNTs are synthesized at 800$^\circ$C for 1 min using Ar and H$_2$ flowing through an ethanol bubbler.

\paragraph*{Photoluminescence Microscopy.}
A homebuilt confocal microscopy system is used to perform PL measurements at room temperature~\cite{Ishii:2015, Jiang:2015, Uda:2016, Higashide:2017}. A wavelength-tunable Ti:sapphire laser is used for excitation after controlling its power and polarization by neutral density filters and a half-wave plate, respectively. The laser beam is focused on the samples using an objective lens with a numerical aperture of 0.8 and a working distance of 3.4~mm. PL is collected through the same objective lens and detected using a liquid-nitrogen-cooled InGaAs diode array attached to a spectrometer. With the laser beam position fixed, we scan the samples mounted on a motorized three-dimensional stage to achieve focusing, sample search over the entire chip, and imaging of the single tubes. All measurements are performed in dry nitrogen to avoid formation of oxygen-induced defects~\cite{Georgi:2008, Yoshikawa:2010}.

\paragraph*{Monte Carlo Simulation.}
We use the same method as in Ref.~\citenum{Ishii:2015}, and modify the excitation profile to reproduce the spatial scanning of the sample. Excitonic processes are evaluated at time intervals $\Delta t\leq 10^{-3}\tau$. For exciton generation, photons are supplied into the system at a rate with a spatial probability distribution given by a 2D Gaussian function. Excitons are generated from the photons on the nanotube with a photon absorption width of 5 nm. Note that a width thicker than the physical width of the nanotube does not affect the following discussion based on the number of generated excitons $g\tau$, and it is used to reduce computational load. Probability for the exciton displacement $s$ due to diffusion is given by the normal distribution $\frac{1}{\sqrt{4\pi D\Delta t}} \exp (-\frac{s^2}{4D\Delta t})$. In addition to the intrinsic decay process that occurs with the probability of $\Delta t/\tau$, the excitons that diffused beyond the tube ends disappear from the system. When two excitons pass by one another, either one of them is eliminated, while the other exciton remains unchanged. The tube length $L$ and the laser radius $r$ are 1.2~$\mu$m and 0.4~$\mu$m, respectively, while the diffusion length $\sqrt{D\tau}=1$~$\mu$m is assumed. 

\begin{acknowledgments}
Work supported in part by JSPS (KAKENHI JP16H05962 and JP17H07359), RIKEN (Incentive Research Project), and MEXT (Nanotechnology Platform). We thank the Advanced Manufacturing Support Team at RIKEN for technical assistance.
\end{acknowledgments}

\end{document}